\documentclass[conference]{IEEEtran}
\IEEEoverridecommandlockouts
\usepackage{cite}
\usepackage{amsmath,amssymb,amsfonts}
\usepackage{algorithmic}
\usepackage{graphicx}
\usepackage{textcomp}
\usepackage{xcolor}
\usepackage{url}
\usepackage{hyperref}
\def\BibTeX{{\rm B\kern-.05em{\sc i\kern-.025em b}\kern-.08em
    T\kern-.1667em\lower.7ex\hbox{E}\kern-.125emX}}
\begin{document}

\title{From Connectivity to Autonomy: The Dawn of Self-Evolving Communication Systems\\
\thanks{This work is funded by the Communications Hub for Empowering Distributed Cloud Computing Applications and Research (CHEDDAR), supported by the EPSRC through UKRI’s Technology Missions Fund.}
}

\author{
    \IEEEauthorblockN{Zeinab Nezami\IEEEauthorrefmark{1}\IEEEauthorrefmark{2}, 
    Syed Danial Ali Shah\IEEEauthorrefmark{1},
    Maryam Hafeez\IEEEauthorrefmark{1},  
    Karim Djemame\IEEEauthorrefmark{2},  
    Syed Ali Raza Zaidi\IEEEauthorrefmark{1}} 
    \IEEEauthorblockA{\IEEEauthorrefmark{1}School of Electronic and Electrical Engineering, University of Leeds, UK}
    \IEEEauthorblockA{\IEEEauthorrefmark{2}School of Computing, University of Leeds, UK}
    \{Z.Nezami, s.shah, M.Hafeez, K.Djemame, S.A.Zaidi\}@leeds.ac.uk

\thanks{}
\thanks{Manuscript received April 19, 2021; revised August 16, 2021.}}

\maketitle

\begin{abstract}

This paper envisions 6G as a self-evolving telecom ecosystem, where AI-driven intelligence enables dynamic adaptation beyond static connectivity. We explore the key enablers of autonomous communication systems, spanning reconfigurable infrastructure, adaptive middleware, and intelligent network functions, alongside multi-agent collaboration for distributed decision-making. We explore how these methodologies align with emerging industrial IoT frameworks, ensuring seamless integration within digital manufacturing processes. Our findings emphasize the potential for improved real-time decision-making, optimizing efficiency, and reducing latency in networked control systems. The discussion addresses ethical challenges, research directions, and standardization efforts, concluding with a technology stack roadmap to guide future developments. By leveraging state-of-the-art 6G network management techniques, this research contributes to the next generation of intelligent automation solutions, bridging the gap between theoretical advancements and real-world industrial applications.
\end{abstract}



Wireless communication has rapidly evolved from basic 2G connectivity to today’s intelligent, data-driven 5G systems. As we move toward the sixth generation (6G), a critical paradigm shift is underway: the emergence of self-evolving communication ecosystems. Unlike traditional adaptive networks that merely react to predefined stimuli, self-evolving systems leverage artificial intelligence (AI) to autonomously perceive, reason, and reconfigure themselves in real time. This marks a structural transformation—enabling communication infrastructures that are scalable, resilient, and context-aware.

Recent research has laid foundational insights into this vision. For example, Self-Evolving Networks (SENs)~\cite{chaoub2022hybrid,qian2024self} have been proposed to manage complexity across integrated terrestrial-aerial-satellite systems through machine learning-driven resource optimization and mobility control~\cite{liang2023code}. Others have introduced multi-layered self-evolving architectures where knowledge is continuously extracted and reapplied across layers to enhance communication system robustness~\cite{chaoub2022hybrid}. Moreover, efforts to create transformative protocol architectures~\cite{cai2022self}, computing frameworks with evolutionary engines~\cite{weyns2023vision}, and neuro-symbolic AI for secure signal processing~\cite{kashikar2024neuro} underscore a growing consensus: future networks must evolve autonomously in response to unforeseen demands and dynamic environments.

Notably, research by Google DeepMind~\cite{hughesposition} underscores the importance of continual innovation and evolution in achieving advanced machine intelligence. However, despite these advancements, key gaps remain. Much of the existing work focuses on architectural concepts or domain-specific applications without fully integrating modular infrastructures, middleware intelligence, and agent-based decision-making in industrial contexts. Furthermore, while simulation results in existing studies demonstrate improved metrics such as signal-to-noise ratio~\cite{kashikar2024neuro} or decision efficiency in IoT~\cite{liu2021framework,lu2023architecture}, few address how such frameworks can be practically deployed in real-world, latency-sensitive environments.

One promising enabler of this shift is the Open-Radio Access Network (O-RAN)\footnote{O-RAN Alliance, \url{https://www.o-ran.org/}, accessed Nov 2024}, which decouples legacy architectures and introduces AI-native control for dynamic, programmable operations. In parallel, the emergence of Large Language Models (LLMs) enhances network intelligence by supporting automation, optimization, and intent-based reconfiguration~\cite{mondal2023llms, bariah2024large}.

This paper addresses the above gaps by presenting a conceptual and architectural vision of a self-evolving telecom ecosystem. Our key contributions are as follows: (i) A  unified technology stack that integrates reconfigurable infrastructure, adaptive middleware, and intelligent network functions to support self-evolving communication systems. (ii) A multi-agent framework for distributed decision-making, enabling real-time autonomy and self-optimization at scale. (iii) A critical discussion of ethical, standardization, and practical implementation challenges for realizing self-evolving communication systems.


By bridging the gap between conceptual innovation and practical deployment, this paper explores how AI–telecom synergy can catalyze the next generation of intelligent networks—creating a globally interconnected digital ecosystem that continuously learns, adapts, and evolves with societal and technological progress.

\section{Autonomous Telecom Ecosystem: Roadmap and Vision}

The transition from traditional telecom networks to self-evolving ecosystems represents a paradigm shift, unfolding in strategic phases. This evolution transforms networks from static infrastructures into autonomous, adaptive systems capable of real-time decision-making and continuous self-improvement, driving efficiency, inclusivity, and societal progress.

The \textit{initial phase} integrates AI and automation into existing telecom infrastructures, enabling AI-driven resource optimization, fault resolution, and traffic management. These enhancements lay the foundation for fully autonomous 6G ecosystems, where AI-powered agents dynamically allocates bandwidth, resolves issues, and optimizes operations without human intervention. This automation improves efficiency and expands connectivity to underserved regions, bridging the digital divide. The \textit{second phase} incorporates Integrated Sensing and Communication (ISAC)\footnote{ISAC, by Robert Baldemair, available at: \url{https://www.ericsson.com/en/blog/2024/6/integrated-sensing-and-communication}} and edge computing~\cite{lin2023pushing}. ISAC enhances situational awareness, allowing networks to sense and respond to real-time conditions, such as adjusting traffic during emergencies. 
Edge computing further reduces latency by enabling localized data processing, ensuring intelligent decision-making. These advancements create highly adaptive networks that support hyper-connectivity in applications such as industrial automation and smart cities, driving economic growth and improving quality of life.

The final phase introduces open-ended learning, where AI autonomously innovates and adapts. Inspired by DeepMind’s work on open-endedness~\cite{hughesposition}, 6G networks will develop novel protocols, adapt to unforeseen scenarios, and implement continuous improvements. Similar to self-driving protein laboratories~\cite{rapp2024self}, which autonomously design and refine experiments, AI-native networks will drive self-optimization. AI scientists~\cite{lu2024ai} further exemplify this potential by autonomously generating hypotheses and analyzing results, reinforcing the vision of self-evolving telecom systems. These systems build upon the Self-X paradigm~\cite{kephart2003vision}, enabling self-healing, self-optimizing networks that proactively adapt, resolve issues, and enhance performance without human intervention. To align with ethical and societal considerations, networks will integrate semantic knowledge representation, allowing them to interpret contexts, prioritize critical services, and ensure equitable access. 

Human-in-the-loop mechanisms will remain crucial for ethical oversight, ensuring decisions align with both technical and social responsibility. By integrating contextual understanding, these systems will dynamically adapt to real-world conditions, optimizing performance across diverse environments. This approach balances external factors (e.g., environmental, economic, and social conditions) with internal network health and performance, fostering a telecom ecosystem that is efficient, fair, and sustainable.
Finally, intent-driven frameworks powered by multimodal LLMs will refine human-machine interactions. Users will specify high-level intents (e.g., ``ensure seamless connectivity for a video conference''), and networks will autonomously reconfigure to meet these needs. These advancements will redefine telecommunications, enabling applications in autonomous transportation, predictive healthcare, and digital equity while fostering new business models and economic growth.
\begin{figure*}[ht]
    \centering
    \includegraphics[clip, trim=0.3cm 6cm 0.3cm 8.5cm,height=14cm,width=\textwidth]{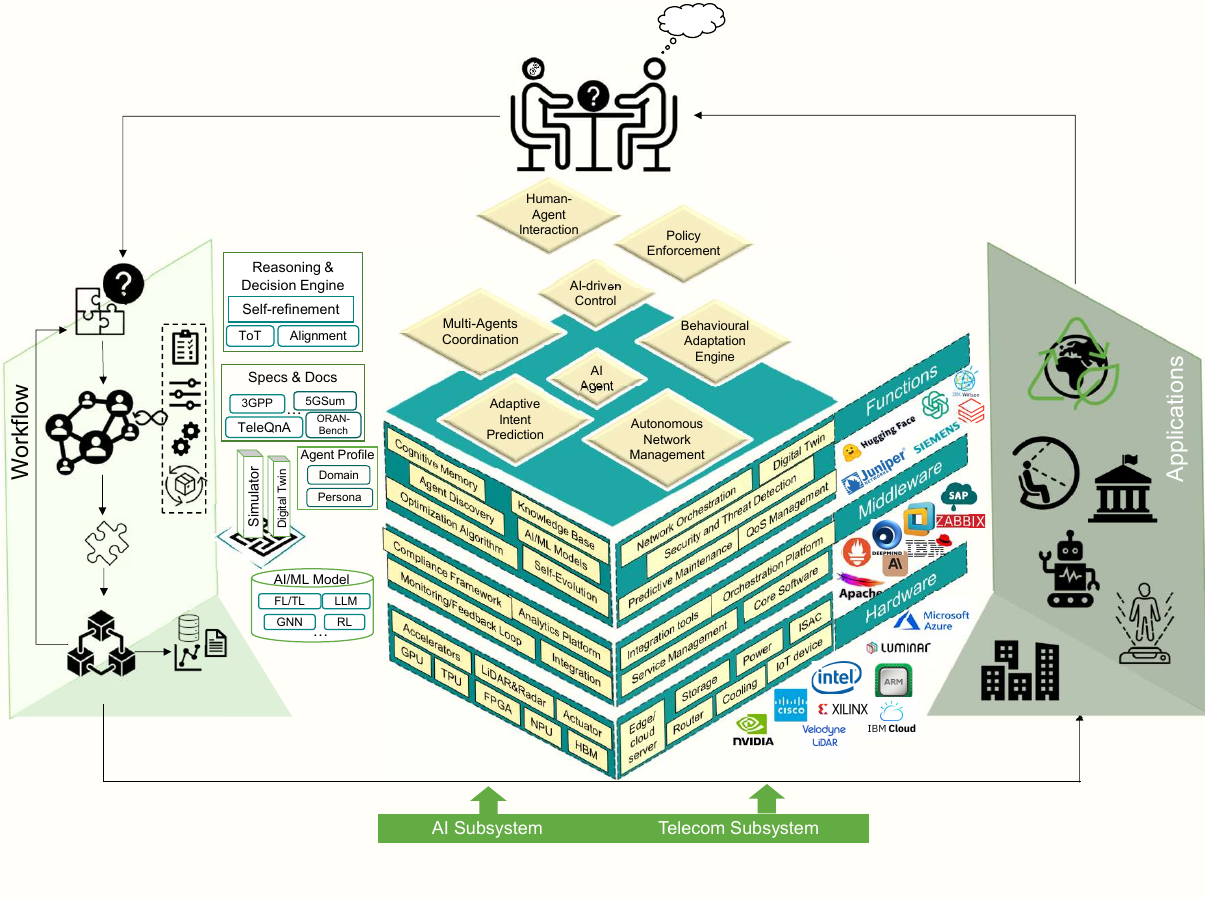}
    \caption{Architecture of AI-6G/ORAN for Self-Evolving Communication Systems}
    \label{fig:architecture}
\end{figure*}

\section{Key Enablers of Self-Evolving Communication Systems}

AI and telecom technologies are deeply intertwined, each serving as both an enabler and a beneficiary of the other. Telecom infrastructure supports AI's distributed computing, connectivity, and real-time data exchange, while AI transforms networks into intelligent, adaptive systems capable of self-evolution. This synergy fosters scalability, autonomy, efficiency, and adaptability, driving innovations that redefine industries and enhance global connectivity.

The multi-layered architecture of autonomous telecom networks, shown in Figure.\ref{fig:architecture}, forms a cohesive framework for advanced communication systems. This figure signifies how Telecom and AI subsystems constitute each layer of the network. At the base, the Hardware Layer provides a reconfigurable infrastructure for seamless data transmission. The Middleware Layer introduces programmability and scalability, enabling networks to dynamically adapt. The Functions Layer consists of modular operations that optimize both AI and telecom capabilities. These modules enhance network intelligence and decision-making. Above them, AI Agents collaborate to continuously optimize network operations, fostering autonomous function, self-evolution, and continuous innovation, transforming telecom systems into intelligent, self-sustaining ecosystems.

\subsection{Hardware Layer: Reconfigurable and Versatile Infrastructure}
The Hardware Layer is fundamental to autonomous telecom ecosystems, offering reconfigurable infrastructure to meet the evolving demands of future communication systems. By integrating traditional telecom components with AI-driven hardware, it supports seamless data transmission, unprecedented compute power, scalability, and autonomous operations. Its adaptability ensures the network evolves with technological advancements and changing user needs, supporting both Human-to-Machine (H2M) and Machine-to-Machine (M2M) interactions in an increasingly interconnected environment.

\textbf{Telecom components} at this layer mainly essential components such as routers, edge servers, and IoT devices that provide reliable connectivity by dynamically allocating resources based on real-time demands. IoT devices and sensors gather contextual data, enabling the network to adapt and respond intelligently. The inclusion of ISAC adds environmental awareness, empowering devices to autonomously optimize operations, improving M2M communication, resource allocation, and safety without human intervention. \textbf{AI components} at this layer enhance the hardware layer's versatility by integrating cognitive capabilities. High Bandwidth Memory (HBM), AI accelerators, Neural Processing Unit (NPU), and Field-Programmable Gate Arrays (FPGAs) provide the computational power for real-time data processing, enabling machine learning models to function efficiently at the edge. These components are designed to evolve with autonomous systems, improving decision-making and adaptability.

The hardware layer also supports H2M interactions through reconfigurable AI hardware, such as the RAN Intelligent Controller (RIC)~\cite{balasubramanian2021ric}, which optimizes network resources for tasks such as video conferencing and real-time gaming. By integrating NLP models and machine learning accelerators, it enables intuitive, personalized user experiences.
Physical Layer AI Agents harness the hardware’s capabilities to optimize parameters such as power control, frequency allocation, and beamforming, enhancing data transmission. These agents manage interference, improve signal quality, and adapt to real-time changes, facilitating efficient spectrum sharing and resource utilization, even in high-demand scenarios. These capabilities leverage the intersection of AI and telecom subsystems at this layer.

\subsection{Middle-ware Layer: Enabling Programmable and Adaptive Networks}

The Middle-ware Layer serves as a bridge between the hardware and AI-driven systems, enabling programmability, scalability, and adaptability across network functions. It allows telecom networks to dynamically respond to evolving demands, fostering continuous innovation and enabling the transition from rigid architectures to self-evolving, intelligent ecosystems.

The \textbf{AI components} at this layer include service management and orchestration for efficient deployment and scaling of AI services, the core AI software layer for real-time decision-making, and data integration frameworks for preparing data for AI models. AI-enabled analytics platforms offer predictive and real-time analysis, while edge AI integration reduces latency. Interoperability frameworks ensure smooth data exchange, and the security and privacy framework guarantees regulatory compliance. Performance monitoring loops enable ongoing optimization, supporting autonomous, adaptive telecom networks.
The \textbf{Telecom components} such as Software-Defined Networking (SDN) and Network Function Virtualization (NFV) provide the backbone for programmable networks, enabling dynamic reconfiguration and optimization. At this layer, platforms such as xApps and rApps offer modular, customizable solutions for tasks such as fault detection and performance optimization. Telemetry pipelines collect and analyze real-time data, while orchestration platforms such as MANO~\cite{9620858} and Kubernetes streamline deployment and scaling of containerized functions. By integrating APIs and telemetry, the system ensures efficient communication and data-driven decision-making, creating a responsive, resilient network environment.

\subsection{Functionality and Operation: The Autonomy Core of Intelligent Networks}

The Functionality and Operation layer serves as the cognitive core of next-generation telecom systems, transforming them into dynamic, self-evolving ecosystems. By integrating advanced AI techniques, this layer ensures continuous learning, real-time adaptation, and intelligent decision-making, enabling 6G networks to drive innovation and optimize performance autonomously.

\textbf{AI components} such as Federated Learning, Transfer Learning, and Reinforcement Learning are key to this process. These models enable the network to learn from distributed data, autonomously adapt to challenges, and optimize operations. Federated Learning ensures privacy through decentralized learning, Transfer Learning accelerates adaptation across domains, and Reinforcement Learning supports self-optimization. These AI techniques form the foundation for self-healing, self-optimizing telecom systems. The Self-Evolving Framework drives continuous learning, allowing AI systems to autonomously generate innovative solutions to dynamic challenges. Open-endedness in AI enables the network to innovate on the fly, while Optimization (e.g., evolutionary) Algorithms facilitate efficient resource allocation, traffic routing, and load balancing. Cognitive Memory stores past decisions, refining future ones, and Retrieval Augmented Generation (RAG) enhances AI agents by providing up-to-date contextual information. Visualization and Reporting tools offer intuitive dashboards for monitoring AI-driven insights. The \textbf{Telecom components} include Network Orchestration, automating and coordinating critical functions such as traffic management, resource allocation, and fault resolution. Quality of Service (QoS) Management prioritizes latency-sensitive services, ensuring optimal performance. Anomaly Detection and Predictive Maintenance ensure network health, with proactive interventions preventing disruptions. Security systems safeguard against threats, while Digital Twins enable network simulation and configuration testing, optimizing real-time performance and planning.

\subsection{Multi-agent Collaborative Telecom System}

The multi-agent collaborative system is key to autonomous network management, where distributed AI agents collaborate to optimize network functionality. These agents, built from lower-layer modules and leveraging advanced AI models, network orchestration, and optimization components, ensure efficient coordination across the network infrastructure.

Each agent autonomously handles tasks such as adaptive intent prediction, anticipating network needs and adjusting operations to maintain optimal service quality and resource utilization. By collaborating as a cohesive multi-agent system, agents make data-driven decisions in real time, improving operational efficiency and reducing manual intervention.
The collaboration between agents is facilitated by a Behavioral Adaptation Engine, enabling continuous evolution of agent actions in response to changing network conditions. This allows for the system to continuously address new challenges and dynamic demands. By ensuring smooth interaction and information sharing among agents, the system can perform complex tasks like fault resolution and traffic optimization in real-time. AI actions are governed by policy enforcement, ensuring compliance with regulatory and ethical standards. Additionally, human-agent interaction modules allow for operator oversight when needed, enhancing network autonomy while maintaining human engagement for critical decisions. This synergy between AI and human oversight strengthens network resilience, adaptability, and efficiency, preparing the system for future challenges.

\subsection{Applications}
Self-evolving communication systems are poised to revolutionize various sectors, especially governance and smart cities. By autonomously managing urban resources such as traffic, energy, and water, these AI-driven networks promote sustainability and contribute to achieving Sustainable Development Goals (SDGs)~\cite{sdgs}. In governance, they enable real-time data analysis and adaptive policy-making, improving public service responsiveness and optimizing management across healthcare, utilities, and other sectors.

Beyond governance, these networks will transform immersive and collaborative experiences, as well as critical infrastructure. In augmented reality (AR), virtual reality (VR), mixed reality (MR), and holography, they ensure seamless, high-quality interactions by minimizing latency and optimizing bandwidth. Critical infrastructure, such as healthcare, transportation, and energy, will benefit from low-latency AI-driven networks that support telemedicine, remote surgeries, and dynamic power grid management. 
Autonomous systems, including self-driving vehicles, drones, and robots, will leverage these networks to adapt and collaborate in real time, boosting safety and efficiency across industries. In summary, self-evolving networks foster smarter, more resilient, and adaptable environments, driving innovation and improving operational efficiency across diverse sectors.

\section{Discussion}

Self-evolving communication systems represent the next frontier of telecom networks, where AI transcends connectivity, addressing societal and technological challenges through autonomy. This section explores their ethical implications, future research directions, standardization needs, and required technological components.

\subsection{Ethical Considerations and Challenges}
The integration of AI into telecom networks presents both transformative potential and significant ethical challenges. Key concerns include bias and fairness, as AI models such as LLMs may propagate biases present in training data. The amplification of biases through feedback loops, particularly with synthetic data, risks exacerbating digital inequality by favoring specific groups or geographies in resource allocation and traffic prioritization. Addressing these risks requires integrating fairness and equity principles, supported by robust bias detection and mitigation techniques. Additionally, privacy concerns arise from the extensive real-time data collection, demanding strict anonymization, encryption, and compliance with regulations such as General Data Protection Regulation (GDPR)~\cite{gdpr} and California Consumer Privacy Act (CCPA)~\cite{ccpa}.

Accountability and transparency are further challenges, particularly when autonomous systems make errors, such as mismanaging network resources. Explainable AI (XAI) frameworks~\cite{brik2024explainable} are critical for ensuring trust and assigning responsibility. Ethical considerations in sectors such as healthcare and transportation are particularly high-stakes, necessitating oversight committees\footnote{AI Safety Institute, available at \url{https://www.aisi.gov.uk/}} to mitigate risks. The societal impact, including potential job market disruptions, calls for proactive upskilling initiatives. Additionally, the increasing sophistication of these systems heightens their vulnerability to cyberattacks, including unauthorized surveillance, requiring stringent security protocols. The high cost of these technologies may also contribute to global inequalities, making equitable access crucial for maximizing societal benefits.

\subsection{Research Directions}
The transition to autonomous networks requires interdisciplinary collaboration across AI, telecommunications, and systems engineering. Key research areas include developing real-time AI models capable of adaptive learning in dynamic environments, utilizing techniques such as reinforcement learning and hybrid AI frameworks for structured reasoning and problem-solving~\cite{mirzadeh2024gsm,fortune2024nextwave}. 
Multi-agent systems for decentralized coordination and intent-driven networking frameworks that allow users to issue high-level directives are also crucial~\cite{vallinder2024cultural}.

Quantum technologies~\cite{marr2024quantumAI} offer opportunities for secure communication and AI enhancement, pushing the limits of encryption and optimization, and interplanetary communications. Advances in edge computing, cloud infrastructure, and federated learning will support low-latency decision-making and maintain data privacy. Research must also address ethical, regulatory, and security challenges through collaboration between researchers, engineers, and policymakers to ensure technical innovations align with societal values, emphasizing sustainability and energy-efficient AI models.

\subsection{Standardization Efforts}

The development of Autonomous Networks requires a forward-thinking approach to standardization, as current frameworks are insufficient for supporting dynamic, AI-driven ecosystems in future communication systems. While organizations such as ITU, IEEE, and 3GPP have laid foundational work, more progress is needed to address the complexities of self-evolving systems.

The ITU-T Focus Group on AI-Native Networks (FG AINN)~\cite{itu_fg_ainn} has made notable contributions in identifying necessary architectural changes, but its efforts must broaden to include protocols for quantum-enhanced AI and decentralized multi-agent collaboration. IEEE’s ``CertifAIEd'' program~\cite{ieee_certifaied} provides a basis for trustworthy AI, but future standards must account for ethical concerns, such as algorithmic accountability in open-ended learning systems and equitable distribution of AI capabilities. The AI-RAN Alliance has focused on AI-driven RAN optimization, but standards for ultra-low latency and advanced spectrum sharing—particularly for holography and real-time immersive applications—must be further developed.

The lack of interoperability between diverse AI systems and communication domains is a major challenge. Future standards need to create universal protocols for heterogeneous AI models, enabling seamless collaboration in dynamic environments. These protocols should support real-time self-organization, adaptive fault management, and secure decision-making. Additionally, governance frameworks should focus on adaptive privacy mechanisms to protect user data as AI technologies evolve.
Standardization should prioritize affordable, scalable models for deploying self-evolving systems, addressing the financial and infrastructure challenges in developing regions to ensure equitable global digital growth.

\subsection{Technology Stack Development}

The realization of autonomous 6G networks hinges on a unified technology stack that combines advanced AI models, such as deep learning, reinforcement learning, and multi-agent systems, with existing technologies for adaptive learning, real-time optimization, self-configuration, and autonomous decision-making. Frameworks such as TensorFlow, PyTorch, and ONNX~\cite{onnx} support the development of AI systems, while edge and cloud computing infrastructures, such as , Google Anthos~\cite{google_anthos}, AWS Greengrass~\cite{aws_greengrass}, and Microsoft Azure IoT~\cite{microsoft_azure_iot}, enable localized and large-scale processing. Communication technologies, including 5G network slicing, massive MIMO, SDN, and AI-driven orchestration platforms such as O-RAN, are further strengthened by testbeds such as Aerial RAN CoLab Over-the-Air, laying the foundation for future 6G networks. O-RAN has transformed traditional cellular networks by enabling open interfaces and multi-vendor interoperability, paving the way for efficient resource utilisation and dynamic network optimization.

In support of this vision, we have implemented and evaluated a dynamic orchestration framework in our prior work~\cite{shah2025interplay}, where the Near Real-Time RIC (NRT-RIC) is extended with a monitoring xApp and an AI-powered orchestrator. The orchestrator, equipped with a Soft Actor-Critic (SAC) reinforcement learning algorithm, dynamically allocates GPU resources between latency-sensitive RAN workloads and generative AI applications. This implementation demonstrates how intelligent resource management can be achieved in real time on shared infrastructure—achieving nearly 99\% service satisfaction for RAN requests while supporting concurrent AI workloads. The system also exemplifies modular integration of AI agents into RAN, validating the architecture presented in this paper.

Security will be addressed through blockchain-based authentication, AI-driven intrusion detection (e.g., IBM QRadar~\cite{chakrabarty2021securing}), and zero-trust architectures. Federated learning platforms (e.g., NVIDIA FLARE~\cite{roth2022nvidia}) will ensure privacy-preserving AI. Real-time feedback loops, knowledge graph integration, and federated learning will enable dynamic adaptation and the incorporation of emerging technologies. This evolving stack forms the backbone of autonomous 6G networks, enhancing scalability, resilience, and intelligence, while our implementation provides a working example of bridging theory and practice.

\section*{ACKNOWLEDGMENT}

This work was funded by the Engineering and Physical Sciences Research Council (EPSRC) – UK Research and Innovation (UKRI) via the Technology Missions Fund (EP/Y037421/1) under project the Communications Hub for Empowering Distributed Cloud Computing Applications and Research (CHEDDAR).

\bibliographystyle{IEEEtran}
\bibliography{references}

\begin{thebibliography}{10}
\providecommand{\url}[1]{#1}
\csname url@samestyle\endcsname
\providecommand{\newblock}{\relax}
\providecommand{\bibinfo}[2]{#2}
\providecommand{\BIBentrySTDinterwordspacing}{\spaceskip=0pt\relax}
\providecommand{\BIBentryALTinterwordstretchfactor}{4}
\providecommand{\BIBentryALTinterwordspacing}{\spaceskip=\fontdimen2\font plus
\BIBentryALTinterwordstretchfactor\fontdimen3\font minus \fontdimen4\font\relax}
\providecommand{\BIBforeignlanguage}[2]{{%
\expandafter\ifx\csname l@#1\endcsname\relax
\typeout{** WARNING: IEEEtran.bst: No hyphenation pattern has been}%
\typeout{** loaded for the language `#1'. Using the pattern for}%
\typeout{** the default language instead.}%
\else
\language=\csname l@#1\endcsname
\fi
#2}}
\providecommand{\BIBdecl}{\relax}
\BIBdecl

\bibitem{chaoub2022hybrid}
A.~Chaoub, A.~Mammel, P.~Martinez-Julia, R.~Chaparadza, M.~Elkotob, L.~Ong, D.~Krishnaswamy, A.~Anttonen, and A.~Dutta, ``Hybrid self-organizing networks: Evolution, standardization trends, and a 6g architecture vision: submitted (is under review) to ieee magazine on standards for evolving and future networks,'' \emph{May 31st}, 2022.

\bibitem{qian2024self}
L.~Qian, P.~Yang, G.~Wu, W.~Tang, and Z.~Chen, ``Self-evolving wireless communications: A novel intelligence trend for 6g and beyond,'' in \emph{2024 International Conference on Future Communications and Networks (FCN)}.\hskip 1em plus 0.5em minus 0.4em\relax IEEE, 2024, pp. 1--6.

\bibitem{liang2023code}
J.~Liang, W.~Huang, F.~Xia, P.~Xu, K.~Hausman, B.~Ichter, P.~Florence, and A.~Zeng, ``Code as policies: Language model programs for embodied control,'' in \emph{2023 IEEE International Conference on Robotics and Automation (ICRA)}.\hskip 1em plus 0.5em minus 0.4em\relax IEEE, 2023, pp. 9493--9500.

\bibitem{cai2022self}
L.~Cai, J.~Pan, W.~Yang, X.~Ren, and X.~Shen, ``Self-evolving and transformative protocol architecture for 6g,'' \emph{IEEE Wireless Communications}, vol.~30, no.~4, pp. 178--186, 2022.

\bibitem{weyns2023vision}
D.~Weyns, T.~B{\"a}ck, R.~Vidal, X.~Yao, and A.~N. Belbachir, ``The vision of self-evolving computing systems,'' \emph{Journal of Integrated Design and Process Science}, vol.~26, no. 3-4, pp. 351--367, 2023.

\bibitem{kashikar2024neuro}
R.~Kashikar, ``Neuro-symbolic ai for self-evolving signal processing in autonomous communication systems,'' in \emph{2024 7th International Conference on Signal Processing and Information Security (ICSPIS)}.\hskip 1em plus 0.5em minus 0.4em\relax IEEE, 2024, pp. 1--6.

\bibitem{hughesposition}
E.~Hughes, M.~D. Dennis, J.~Parker-Holder, F.~Behbahani, A.~Mavalankar, Y.~Shi, T.~Schaul, and T.~Rockt{\"a}schel, ``Position: Open-endedness is essential for artificial superhuman intelligence,'' in \emph{Forty-first International Conference on Machine Learning}, 2024.

\bibitem{liu2021framework}
B.~Liu, J.~Luo, and X.~Su, ``The framework of 6g self-evolving networks and the decision-making scheme for massive iot,'' \emph{Applied Sciences}, vol.~11, no.~19, p. 9353, 2021.

\bibitem{lu2023architecture}
L.~Lu, C.~Liu, C.~Zhang, Z.~Hu, S.~Lin, Z.~Liu, M.~Zhang, X.~Liu, and J.~Chen, ``Architecture for self-evolution of 6g core network based on intelligent decision making,'' \emph{Electronics}, vol.~12, no.~15, p. 3255, 2023.

\bibitem{mondal2023llms}
R.~Mondal, A.~Tang, R.~Beckett, T.~Millstein, and G.~Varghese, ``What do llms need to synthesize correct router configurations?'' in \emph{Proceedings of the 22nd ACM Workshop on Hot Topics in Networks}, 2023, pp. 189--195.

\bibitem{bariah2024large}
L.~Bariah, Q.~Zhao, H.~Zou, Y.~Tian, F.~Bader, and M.~Debbah, ``Large generative ai models for telecom: The next big thing?'' \emph{IEEE Communications Magazine}, 2024.

\bibitem{lin2023pushing}
Z.~Lin, G.~Qu, Q.~Chen, X.~Chen, Z.~Chen, and K.~Huang, ``Pushing large language models to the 6g edge: Vision, challenges, and opportunities,'' \emph{arXiv preprint arXiv:2309.16739}, 2023.

\bibitem{rapp2024self}
J.~T. Rapp, B.~J. Bremer, and P.~A. Romero, ``Self-driving laboratories to autonomously navigate the protein fitness landscape,'' \emph{Nature chemical engineering}, vol.~1, no.~1, pp. 97--107, 2024.

\bibitem{lu2024ai}
C.~Lu, C.~Lu, R.~T. Lange, J.~Foerster, J.~Clune, and D.~Ha, ``The ai scientist: Towards fully automated open-ended scientific discovery,'' \emph{arXiv preprint arXiv:2408.06292}, 2024.

\bibitem{kephart2003vision}
J.~O. Kephart and D.~M. Chess, ``The vision of autonomic computing,'' \emph{Computer}, vol.~36, no.~1, pp. 41--50, 2003.

\bibitem{balasubramanian2021ric}
B.~Balasubramanian, E.~S. Daniels, M.~Hiltunen, R.~Jana, K.~Joshi, R.~Sivaraj, T.~X. Tran, and C.~Wang, ``Ric: A ran intelligent controller platform for ai-enabled cellular networks,'' \emph{IEEE Internet Computing}, vol.~25, no.~2, pp. 7--17, 2021.

\bibitem{9620858}
J.~Lee and Y.~Kim, ``A design of mano system for cloud native infrastructure,'' in \emph{2021 International Conference on Information and Communication Technology Convergence (ICTC)}, 2021, pp. 1336--1339.

\bibitem{sdgs}
\BIBentryALTinterwordspacing
{United Nations}, ``Sustainable development goals,'' 2015, accessed: 2025-04-05. [Online]. Available: \url{https://sdgs.un.org/goals}
\BIBentrySTDinterwordspacing

\bibitem{gdpr}
\BIBentryALTinterwordspacing
{European Union}, ``General data protection regulation (gdpr),'' 2016, accessed: 2025-04-05. [Online]. Available: \url{https://www.consilium.europa.eu/en/policies/data-protection-regulation/}
\BIBentrySTDinterwordspacing

\bibitem{ccpa}
\BIBentryALTinterwordspacing
{State of California}, ``California consumer privacy act (ccpa),'' 2018, accessed: 2025-04-05. [Online]. Available: \url{https://oag.ca.gov/privacy/ccpa}
\BIBentrySTDinterwordspacing

\bibitem{brik2024explainable}
B.~Brik, H.~Chergui, L.~Zanzi, F.~Devoti, A.~Ksentini, M.~S. Siddiqui, X.~Costa-P{\`e}rez, and C.~Verikoukis, ``Explainable ai in 6g o-ran: A tutorial and survey on architecture, use cases, challenges, and future research,'' \emph{IEEE Communications Surveys \& Tutorials}, 2024.

\bibitem{mirzadeh2024gsm}
I.~Mirzadeh, K.~Alizadeh, H.~Shahrokhi, O.~Tuzel, S.~Bengio, and M.~Farajtabar, ``Gsm-symbolic: Understanding the limitations of mathematical reasoning in large language models,'' \emph{arXiv preprint arXiv:2410.05229}, 2024.

\bibitem{fortune2024nextwave}
\BIBentryALTinterwordspacing
V.~Wadhwa. (2024, October) The next wave of ai won’t be driven by llms. here’s what investors should focus on instead. Accessed: 2025-01-02. [Online]. Available: \url{https://fortune.com/2024/10/18/next-wave-ai-llms-investor-focus-tech/}
\BIBentrySTDinterwordspacing

\bibitem{vallinder2024cultural}
A.~Vallinder and E.~Hughes, ``Cultural evolution of cooperation among llm agents,'' \emph{arXiv preprint arXiv:2412.10270}, 2024.

\bibitem{marr2024quantumAI}
\BIBentryALTinterwordspacing
B.~Marr, ``The next breakthrough in artificial intelligence: How quantum ai will reshape our world,'' \emph{Forbes}, October 8 2024, accessed: 2025-01-28. [Online]. Available: \url{https://www.forbes.com/sites/bernardmarr/2024/10/08/the-next-breakthrough-in-artificial-intelligence-how-quantum-ai-will-reshape-our-world/}
\BIBentrySTDinterwordspacing

\bibitem{itu_fg_ainn}
\BIBentryALTinterwordspacing
{International Telecommunication Union}, ``Focus group on artificial intelligence native for telecommunication networks (fg-ainn),'' 2024, accessed: 2025-04-05. [Online]. Available: \url{https://www.itu.int/en/ITU-T/focusgroups/ainn/Pages/default.aspx}
\BIBentrySTDinterwordspacing

\bibitem{ieee_certifaied}
\BIBentryALTinterwordspacing
{IEEE}, ``Ai ethics certification,'' 2025, accessed: 2025-04-05. [Online]. Available: \url{https://engagestandards.ieee.org/ieeecertifaied.html}
\BIBentrySTDinterwordspacing

\bibitem{onnx}
\BIBentryALTinterwordspacing
{ONNX}, ``Open neural network exchange (onnx),'' 2025, accessed: 2025-04-05. [Online]. Available: \url{https://onnx.ai/}
\BIBentrySTDinterwordspacing

\bibitem{google_anthos}
\BIBentryALTinterwordspacing
{Google Cloud}, ``Anthos,'' 2025, accessed: 2025-04-05. [Online]. Available: \url{https://cloud.google.com/anthos}
\BIBentrySTDinterwordspacing

\bibitem{aws_greengrass}
\BIBentryALTinterwordspacing
{Amazon Web Services}, ``Aws iot greengrass,'' 2025, accessed: 2025-04-05. [Online]. Available: \url{https://aws.amazon.com/greengrass/}
\BIBentrySTDinterwordspacing

\bibitem{microsoft_azure_iot}
\BIBentryALTinterwordspacing
{Microsoft}, ``Microsoft azure iot,'' 2025, accessed: 2025-04-05. [Online]. Available: \url{https://azure.microsoft.com/en-us/solutions/iot/}
\BIBentrySTDinterwordspacing

\bibitem{shah2025interplay}
S.~D.~A. Shah, Z.~Nezami, M.~Hafeez, and S.~A.~R. Zaidi, ``The interplay of ai-and-ran: Dynamic resource allocation for converged 6g platform,'' \emph{arXiv preprint arXiv:2503.07420}, 2025.

\bibitem{chakrabarty2021securing}
B.~Chakrabarty, S.~R. Patil, S.~Shingornikar, A.~Kothekar, P.~Mujumdar, S.~Raut, D.~Ukirde \emph{et~al.}, \emph{Securing Data on Threat Detection by Using IBM Spectrum Scale and IBM QRadar: An Enhanced Cyber Resiliency Solution}.\hskip 1em plus 0.5em minus 0.4em\relax IBM Redbooks, 2021.

\bibitem{roth2022nvidia}
H.~R. Roth, Y.~Cheng, Y.~Wen, I.~Yang, Z.~Xu, Y.-T. Hsieh, K.~Kersten, A.~Harouni, C.~Zhao, K.~Lu \emph{et~al.}, ``Nvidia flare: Federated learning from simulation to real-world,'' \emph{arXiv preprint arXiv:2210.13291}, 2022.

\end{thebibliography}

\end{document}